\documentclass[a4paper,12pt]{article}
\usepackage{tabularx,hhline,multirow,amsmath,amssymb}

\renewcommand{\bf}[1]{\textbf{#1}}

\renewcommand{\sl}[1]{\textsl{#1}}

\newcommand{\mcal}{\mathcal}

\newcommand{\mbf}{\mathbf}

\newcommand{\R}{\ensuremath{\mathbb{R}}}

\newcommand{\C}{\ensuremath{\mathbb{C}\,}}
\newcommand{\Z}{\ensuremath{\mathbb{Z}}}

\newcommand{\co}{\textrm{\lq\lq}}

\newcommand{\Om}{\ensuremath{\Omega}}
\newcommand{\om}{\ensuremath{\omega}}
\newcommand{\TR}{\ensuremath{\text{Tr}}}
\newcommand{\onh}{\ensuremath{\overline{\mcal{NH}}}}
\newcommand{\ONH}{\ensuremath{\overline{NH}}}

\newcommand{\brho}{\ensuremath{\boldsymbol{\rho}}}
\newcommand{\bvarkappa}{\ensuremath{\pmb{\varkappa}}}

\def\be#1\ee{\begin{equation}#1\end{equation}}
\def\benn#1\eenn{\begin{equation*}#1\end{equation*}}
\newcommand{\ba}{\begin{array}}
\newcommand{\ea}{\end{array}}
\newcommand{\bea}{\begin{eqnarray}}
\newcommand{\eea}{\end{eqnarray}}

\begin{document}

\thispagestyle{empty}

\begin{center}
{\large{\bf{Classical Systems and Representations of (2+1) Newton--Hooke  
Symmetries}}} 
\end{center}

\vspace{1cm}
 
\begin{center} 
Oscar Arratia $^{1}$, Miguel A. Mart¡n $^{1}$ and 
Mariano A. del Olmo $^{2}$
\end{center}

\begin{center}
\emph{$^{1}$ Departamento de Matem tica Aplicada a la
Ingenier¡a,  \\
Universidad de Valladolid. E-47011, Valladolid, Spain.\\
E. mail: oscarr$@$wmatem.eis.uva.es, migmar$@$wmatem.eis.uva.es}
\end{center}

\begin{center}
\emph{$^{2}$ Departamento de F¡sica Te¢rica,\\
Universidad de Valladolid. E-47011, Valladolid, Spain.\\ 
E. mail: olmo$@$cpd.uva.es}
\end{center}

\vspace{1cm}

\begin{center}
\today
\end{center}

\begin{abstract}
A study of the Newton--Hooke kinematical groups in $(2+1)$ dimensions is 
presented revealing the physical interest of these symmetries. A complete 
classification of both classical and quantum elementary systems is achieved 
by explicit computation of coadjoint orbits and unitary irreducible 
representations of extended groups. In addition, we present an application 
example of quantization \emph{… la Moyal} of a classical system using the 
Stratonovich--Weyl correspondence and also give some ideas about a second 
central extension, which did not appear in the (3+1)--dimensional case. 
\end{abstract}

PACS: 02.20.QS, 03.65.BZ


\section{Introduction}
The Newton--Hooke groups, $NH_{\pm}$, were introduced some years ago by 
Bacry and L'vy--Leblond \cite{BL68} as Newton groups. Later, Derome and 
Dubois \cite{DD72} carried out a complete study of these groups in the 
(3+1)--dimensional (D) case giving them the name of Hooke groups. These 
groups appeared while looking for all the possible realizations of the 
relativity principle. So, they are kinematical groups, such as Galilei and 
Poincar' ones. They can be considered non-relativistic cosmological groups 
characterized by a temporal constant $\tau$ which determines the 
characteristic time of the universe stable under the group action. 

It is well--known in Physics the fact that restriction to one spatial 
dimension implies a great simplification in mathematical models to be used 
which entails a substantial decrease in calculations. However, this is not 
the case when one consider 2D models, in which case the symmetry of the 
systems results in a wide variety of concepts. 

Classical and quantum systems associated with (1+1)D kinematical groups 
have been studied in relation with quantization problem 
\cite{GMN91}--\cite{AO97b}. Concerning (2+1)D kinematical groups, Ref. 
\cite{BGO92} follows the same patterns for the Galilei group and Refs.
\cite{Bos95}--\cite{Gri96b} are only focused on representation theory for 
Galilei or Poincar' groups. 

In this context, the intrinsic interest of Newton--Hooke $(2+1)$ groups 
(related with the harmonic oscillator and systems in expansion as we will 
see later on) is enlarged with a rich structure of central extensions and a 
wide set of different classical (coadjoint orbits) and quantum (irreducible 
representations) elementary systems. 

In the framework of In"n--Wigner contractions \cite{IW53}, the groups we 
deal with here correspond to space--velocity kind contractions of the de 
Sitter or Anti--de Sitter groups, $dS_{\pm}$, and leads to Galilei group 
through a space--time contraction. This property is important for the sake 
of interpretation as we will see in section \ref{sec:remarks}. In these 
sense, $NH_{\pm}$ groups can be considered as non-relativistic cosmological 
kinematical groups. On the other hand, the study of Hooke cosmology reveals 
that the main difference between Newton--Hooke and Galilei groups is the 
existence of a harmonic oscillator ($NH_-$) or expansion ($NH_+$) potential  
which make these groups can be considered as dynamical groups of galilean 
systems. Characteristic time $\tau$ can then be interpreted in terms of the 
inverse of Hubble's constant \cite{DD72} for the expanding universe or 
associated to the \co period" for the oscillating case. 

The paper is organized as follows. In section \ref{sec:nh21} we summarize 
the basic definitions of the $NH(2+1)$ groups as well as their 
corresponding Lie algebras  and central extensions. Section \ref{sec:orb} 
is devoted to the construction and classification of classical elementary 
systems associated with these groups via the coadjoint action and it 
presents a simple example of physical interpretation. In section 
\ref{sec:qes} we develop the theory of representations of these Lie groups 
in order to get the quantum elementary systems and their classification. In 
section \ref{sec:quant} we give to the  reader the basic references 
concerning Moyal's quantization by means of the Stratonovich--Weyl 
correspondence and present an example of application for elementary systems 
obtained in previous sections. Finally, section \ref{sec:remarks} is 
intended to discuss some results about extensions and expose our 
conclusions. 
 
\section{The (2+1) Newton--Hooke Group}\label{sec:nh21}
We will focus our attention on $NH_-(2+1)$, the oscillating group denoted 
$NH$ henceforth, the study  for $NH_+$ is similar and can be achieved by 
replacing trigonometric functions by hyperbolic ones in group law. Thus, 
$NH$ can be considered the group of transformations of space--time acting 
on the point of coordinates $(t,\mbf{x})$ by 
\be\label{eq:transfnh21}
(t',\mbf{x}')=g(t,\mbf{x})=\bigl(t+b,\; \mbf{x}^{\phi}
         +\mbf{v}\tau\sin\frac{t}{\tau}+\mbf{a}\cos\frac{t}{\tau}\bigr),
\ee
where $\mbf{x}^{\phi}=R(\phi)\mbf{x}$ stands for the two--dimensional 
vector $\mbf{x}$ rotated an angle $\phi$ $\big(R(\phi)\in SO(2)\big)$. 
Explicitly, $\mbf{x}^{\phi}=\bigl(\begin{smallmatrix} 
\cos\phi&-\sin\phi\\ 
\sin\phi&\cos\phi\end{smallmatrix}\bigr)
\bigl(\begin{smallmatrix}x_1\\x_2\end{smallmatrix}\bigr)$. 

The group elements are parametrized by 
\be
g=(b,\mbf{a},\mbf{v},\phi),\qquad b,\phi\in\R;\;\mbf{a},\mbf{v}\in\R^2,
\ee
$\mbf{a}$, $b$ being the parameters of space--time translations, $\mbf{v}$ 
that of pure $NH$ inertial transformations (boosts) and $\phi$ that of 
rotations.

The constant $\tau$ symbolizes the proper or characteristic time we 
mentioned before \cite{DD72} and is not reabsorbed in time units for the 
sake of physical interpretation. 

The corresponding Lie algebra, $\mcal{NH}$, has dimension six and is 
spanned by the infinitesimal generators of time translations, $H$, space 
translations, $\mbf{P}\equiv(P_1,P_2)$, boosts, $\mbf{K}\equiv(K_1,K_2)$, 
and rotations around an axis perpendicular to the plane, $J$. Their 
non--vanishing commutation relations are 
\be\ba{lrll}
[J,P_i]=\epsilon_{ij}P_j, &\qquad& [P_i,H]=\frac{-1}{\tau^2}K_i,&\\[.4cm]
[J,K_i]=\epsilon_{ij}K_j, &\qquad& [K_i,H]=P_i.&\qquad (i,j=1,2), \ea
\ee 
where $\epsilon_{ij}$ is the totally skewsymmetric tensor.

The algebra $\mcal{NH}$ admits a maximal nontrivial central extension by 
$\R^3$, associated to the following new non--zero Lie brackets 
\be\ba{lrlrl}
[P_1,P_2]=\frac{1}{\tau^2}F,  &\qquad &[P_1,K_1]=-M, 
                                        &\qquad& [H,J]=L,\\[.3cm]
[K_1,K_2]=F,  &\qquad &[P_2,K_2]=-M.& & \ea 
\ee
However, we can neglect the extension associated to $L$ if we realize that 
it is not a group extension even though it is an algebra extension 
\cite{Bar54}. This fact relies on the existence of an infinity to one 
homomorphism relating $SO(2)$ and its universal enveloping group $\R$, 
$(SO(2)\simeq\R/\Z)$. The extended algebra we will consider is therefore 
eight dimensional. 

Since central generators acts trivially on the space--time, the extended 
group acts on $(t,\bf{x})$ in the same way indicated by 
\eqref{eq:transfnh21}. The group law for the extended group \ONH can be 
expressed by 
\be\begin{split}
\bar{g}'\bar{g}&=(\alpha',\theta',b',\mbf{a}',\mbf{v}',\phi')
                 (\alpha,\theta,b,\mbf{a},\mbf{v},\phi)\\[.3cm]
&\equiv (e^{\alpha' F}e^{\theta'M}e^{b'H}e^{\mbf{a}'\mbf{P}}
                                        e^{\mbf{v}'\mbf{K}}e^{\phi'J})
                 (e^{\alpha F}e^{\theta M}e^{bH}e^{\mbf{a}\mbf{P}}
                              e^{\mbf{v}\mbf{K}}e^{\phi J})\\[.3cm]
&=\Bigl(\alpha'+\alpha+\frac{1}{2\tau^2}(\mbf{a}'\cos\frac{b}{\tau}+
        \tau\mbf{v}'\sin\frac{b}{\tau})\times \mbf{a}^{\phi'}+
        \frac{1}{2}(\frac{-\mbf{a}'}{\tau}\sin\frac{b}{\tau}+
        \mbf{v}'\cos\frac{b}{\tau})\times\mbf{v}^{\phi'},\\[.3cm]
&\hspace{.6cm}\theta'+\theta+\frac{1}{2}(\tau\mbf{v}'{}^2-\frac{\mbf{a}'{}^2}
        {\tau})\sin\frac{b}{\tau}\cos\frac{b}{\tau}-
        \mbf{a}'\mbf{v}'\sin^2\frac{b}{\tau}+\mbf{v}'\mbf{a}^{\phi'}
        \cos\frac{b}{\tau}-\frac{\mbf{a}'\mbf{a}^{\phi'}}{\tau}
        \sin\frac{b}{\tau},\\[.3cm]
&\hspace{.6cm}b'+b,\;\mbf{a}'\cos\frac{b}{\tau}+\mbf{v}'\tau\sin
        \frac{b}{\tau}+\mbf{a}^{\phi'},\;\mbf{v}'\cos\frac{b}{\tau}-
        \frac{\mbf{a}'}{\tau}\sin\frac{b}{\tau}+\mbf{v}^{\phi'},
        \;\phi'+\phi\Bigr).\end{split}
\ee 
The inverse of an element $g$ takes the form 
\be\begin{split}
\bar{g}^{-1}&=(\alpha,\theta,b,\mbf{a},\mbf{v},\phi)^{-1}\\[.3cm]
&=\Bigl(-\alpha,\;-\theta-\frac{1}{2}\bigl(\tau\mbf{v}^2-\frac{\mbf{a}^2}{\tau}\bigr) 
        \sin\frac{b}{\tau}\cos\frac{b}{\tau}
        +\mbf{a}\mbf{v}\cos^2\frac{b}{\tau},\\[.3cm] 
&\hspace{.9cm}-b,\;\bigl(\tau\mbf{v}\sin\frac{b}{\tau}-\mbf{a}
        \cos\frac{b}{\tau}\bigr)^{-\phi},\;\bigl(-\mbf{v}\cos\frac{b}{\tau}-
        \frac{\mbf{a}}{\tau}\sin\frac{b}{\tau}\bigr)^{-\phi},\,\,-\phi\Bigr).
\end{split}
\ee
\section{Classical elementary systems of $\ONH$}\label{sec:orb}
The obtention of classical elementary systems with $NH$ symmetry is 
achieved by computing the coadjoint orbits of the extended group. The 
coadjoint action of a generic element 
$g=(\alpha,\theta,b,\mbf{a},\mbf{v},\phi)\in\ONH$ on a point in 
$\onh^{\phantom{,}*}$ , the dual space of the Lie algebra $\onh$, with 
coordinates $(f,m,h,\mbf{p},\mbf{k},j)$ in a dual basis of the basis 
$\{F,M,H,\mbf{P},\mbf{K},J\}$ of $\onh$ (and in this order) can be 
expressed in a compact way as 
\be\label{eq:coadnh21}
\begin{split}
f'&=f,\\[.2cm] m'&=m,\\[.2cm] 
\mbf{p}'&=\cos\frac{b}{\tau}\mbf{p}^{\phi}
          -\frac{1}{\tau}\sin\frac{b}{\tau}\mbf{k}^{\phi}
          -\frac{f}{\tau^2}(\cos\frac{b}{\tau}\mbf{a}^{\pi/2}
          -\tau\sin\frac{b}{\tau}\mbf{v}^{\pi/2})
          -\frac{m}{\tau}(\tau\cos\frac{b}{\tau}\mbf{v}
          +\sin\frac{b}{\tau}\mbf{a}), \\[.3cm]
\mbf{k}'&=\tau\sin\frac{b}{\tau}\mbf{p}^{\phi}
          +\cos\frac{b}{\tau}\mbf{k}^{\phi}
          -\frac{f}{\tau}(\tau\cos\frac{b}{\tau}\mbf{v}^{\pi/2}
          +\sin\frac{b}{\tau}\mbf{a}^{\pi/2})
          +m(\cos\frac{b}{\tau}\mbf{a}
          -\tau\sin\frac{b}{\tau}\mbf{v}), \\[.2cm]
h'&=h-\mbf{v}\mbf{p}^{\phi}+\frac{1}{\tau^2}\mbf{a}\mbf{k}^{\phi}
    +\frac{f}{\tau^2}\mbf{v}\mbf{a}^{\pi/2}
    +\frac{m}{2}(\frac{\mbf{a}^2}{\tau^2}+\mbf{v}^2), \\[.2cm]
j'&=j+\mbf{a}^{\pi/2}\mbf{p}^{\phi}+\mbf{v}^{\pi/2}\mbf{k}^{\phi}
    -\frac{f}{2}(\frac{\mbf{a}^2}{\tau^2}+\mbf{v}^2)
    -m\mbf{v}\mbf{a}^{\pi/2}, \end{split}
\ee

\noindent where $\mbf{p},\,\mbf{k}\in\R^2,\;\;h,\,j\in\R$, and we 
have made use of the following conventions
\be\label{eq:notaciondosvec}
\begin{split}
\mbf{u}\cdot\mbf{v}\equiv\mbf{u}\mbf{v}&=u_1v_1+u_2v_2,\\[.5cm] 
|\mbf{u}|^2\equiv\mbf{u}^2&=u_1^2+u_2^2,\\[.5cm]
 \mbf{u}\times\mbf{v}&=u_1v_2-u_2v_1.\end{split}
\ee 

This action splits $\onh^{\phantom{,}*}$ into several orbits, which can be 
classified according to its invariants. We present here a complete 
classification of the orbits, labeled by the values of the trivial 
invariants $(f,m)$, which corresponds to the extensions, and the 
non--trivial invariants ($C_i$), obtained from  
\eqref{eq:coadnh21} or by using the kernel of the associated Kirillov 
two--form on $\onh^{\phantom{,}*}$ \cite{Nzo88}. In addition, we give the 
dimension of the orbits between brackets. 

\begin{subequations}
\noindent
(\sl{1}) \bf{Non--vanishing extensions $(f,m\neq 0)$:}
\begin{align}
\label{eq:orbnh21a}\blacktriangleright &\quad 
(f\neq \pm m\tau)\left\{ \ba{l}
   C_1=\mbf{p}^2+\frac{\mbf{k}^2}{\tau^2}-2mh+\frac{2}{\tau^2}f j\\[.3cm]
   C_2=\mbf{p}\times\mbf{k}+mj-f h \ea \right\} \quad[4D].\\[.4cm]
\label{eq:orbnh21b}\blacktriangleright &\quad 
(f=m\tau)\left\{ \ba{l}
   0<C_3=\mbf{p}^2+\frac{\mbf{k}^2}{\tau^2}
                        -\frac{2}{\tau}(\mbf{p}\times\mbf{k})\\[.3cm]
   C_4=\mbf{p}^2+\frac{\mbf{k}^2}{\tau^2}
        +\frac{2}{\tau}(\mbf{p}\times\mbf{k})
        -\frac{4f}{\tau}(h-\frac{j}{\tau}) \ea \right\} \quad[4D].\\[.4cm]
\label{eq:orbnh21c}\blacktriangleright &\quad
(f=-m\tau)\left\{ \ba{l}
   0<C_3'=\mbf{p}^2+\frac{\mbf{k}^2}{\tau^2}
                        +\frac{2}{\tau}(\mbf{p}\times\mbf{k})\\[.3cm]
   C_4'=\mbf{p}^2+\frac{\mbf{k}^2}{\tau^2}
        -\frac{2}{\tau}(\mbf{p}\times\mbf{k})
        +\frac{4f}{\tau}(h+\frac{j}{\tau}) \ea \right\} \quad[4D].\\[.4cm]
\label{eq:orbnh21d}\blacktriangleright &\quad
(f=m\tau)\qquad C_3=0,\quad C_4,\quad 
                                C_5=h+\frac{j}{\tau},\quad[2D].\\[.4cm]
\label{eq:orbnh21e}\blacktriangleright &\quad
(f=-m\tau)\qquad C_3'=0,\quad C_4',\quad 
                                C_5'=h-\frac{j}{\tau},\quad[2D].
\end{align}
(\sl{2}) \bf{One null extension:}
\begin{align}
\label{eq:orbnh21f}\blacktriangleright &\quad 
(f=0,m\neq 0)\left\{ \ba{l}
   C_1=\mbf{p}^2+\frac{\mbf{k}^2}{\tau^2}-2mh \\[.3cm]
   C_2=\mbf{p}\times\mbf{k}+mj \ea \right\} \quad[4D].\\[.4cm]
\label{eq:orbnh21g}\blacktriangleright &\quad 
(f\neq 0,m=0)\left\{ \ba{l}
   C_1=\mbf{p}^2+\frac{\mbf{k}^2}{\tau^2}+\frac{2}{\tau^2}f j\\[.3cm]
   C_2=\mbf{p}\times\mbf{k}-f h  \ea \right\} \quad[4D].
\end{align}
(\sl{3}) \bf{Null extensions $(f,m=0)$:}
\begin{align}
\label{eq:orbnh21h}\blacktriangleright &\quad 
(|C_2|<\frac{\tau}{2}C_1)\left\{ \ba{l}
   0<C_1=\mbf{p}^2+\frac{\mbf{k}^2}{\tau^2}\\[.3cm]
   C_2=\mbf{p}\times\mbf{k} \ea \right\} \quad[4D].\\[.4cm]
\label{eq:orbnh21i}\blacktriangleright &\quad 
(C_2=\frac{\tau}{2}C_1)\left\{ \ba{l}
   0<C_1=\mbf{p}^2+\frac{\mbf{k}^2}{\tau^2}\\[.3cm]
   \mbf{k}-\tau\mbf{p}^{\pi/2}=\mbf{0}\\[.3cm]
   C_5=h+\frac{j}{\tau}\ea \right\} \quad[2D].\\[.4cm]
\label{eq:orbnh21j}\blacktriangleright &\quad 
(C_2=\frac{-\tau}{2}C_1)\left\{ \ba{l}
   0<C_1=\mbf{p}^2+\frac{\mbf{k}^2}{\tau^2}\\[.3cm]
   \mbf{k}+\tau\mbf{p}^{\pi/2}=\mbf{0}\\[.3cm]
   C_5'=h-\frac{j}{\tau}\ea \right\} \quad[2D].\\[.4cm]
\label{eq:orbnh21k}\blacktriangleright &\quad 
(C_1=C_2=0)\qquad \mbf{p}=\mbf{k}=\mbf{0},
                \quad h,\quad j,\quad\text{(points)}\;[0D].
\end{align}
\end{subequations}

In the preceding classification we find 4D orbits for non--vanishing values 
of extensions as well as for null values of one of them or both, although 
only orbits labeled by \eqref{eq:orbnh21a}, 
\eqref{eq:orbnh21f} and \eqref{eq:orbnh21g} are spaces diffeomorphic to 
$\R^4$. We must note here the change in the geometry of the orbit when the 
parameters of extensions are related by the equality $f=\pm m\tau$, a fact 
that does not occurs for other (2+1) groups (see \cite{BGO92,Mar98}). 
Orbits \eqref{eq:orbnh21b} and \eqref{eq:orbnh21c} are diffeomorphic to 
$\R^3\times S^1$, while orbits \eqref{eq:orbnh21d} and \eqref{eq:orbnh21e} 
are to $\R^2$. Concerning the orbits corresponding to vanishing extensions, 
\eqref{eq:orbnh21h} is diffeomorphic to $\R^2\times S^1\times S^1$ while 
\eqref{eq:orbnh21i} and \eqref{eq:orbnh21j} are to $\R\times S^1$. 

We can give a dynamical interpretation of the above systems considering 
$b(\equiv t)$ as the parameter of time evolution and using the techniques 
exposed in Ref. \cite{Nzo88} for the Galilei group. As an example, we study 
the dynamics of a system characterized by the pair $(f=0,m\neq 0)$. Later 
on, we will see that this restriction is not very significant from a 
physical point of view when trying to assign a meaning to the central 
extension defined by the parameter $f$. 

So, let us consider an orbit of type \eqref{eq:orbnh21f} denoted by 
$O_{C_1,C_2}^m$. A set of canonical coordinates on the orbit (in the sense 
that their Poisson brackets verify 
$\{q_i,p_j\}=\delta_{ij},\;\{q_1,q_2\}=\{p_1,p_2\}=0$) is determined by 
$\mbf{q}:=\mbf{k}/m$ and $\mbf{p}$. From \eqref{eq:coadnh21} we can write 
the time evolution of these coordinates as 
\be\begin{split}
\mbf{q}(t)&=\frac{\mbf{k}(t)}{m}=\frac{\mbf{p}(0)}{m}\tau\sin\frac{t}{\tau}
                +\mbf{q}(0)\cos\frac{t}{\tau},\\[.3cm]
\mbf{p}(t)&=\mbf{p}(0)\cos\frac{t}{\tau}
                -\frac{m}{\tau}\mbf{q}(0)\sin\frac{t}{\tau}.
\end{split}
\ee 
From here we deduce that evolution equations for the system are 
\be
\frac{d\mbf{q}}{dt}=\frac{\mbf{p}}{m},\qquad
\frac{d\mbf{p}}{dt}=\frac{-m}{\tau^2}\mbf{q},
\ee
and therefore, the hamiltonian function is 
\be
H(\mbf{q},\mbf{p})=\frac{\mbf{p}^2}{2m}+\frac{m}{2\tau^2}\mbf{q}^2+C, 
\ee
which coincides with the value of $h$ obtained from the invariant $C_1$ of 
the orbit and represents the energy of a 2D harmonic oscillator with 
frequency $\om=1/\tau$. Invariants $C_1$ and $C_2$ are interpreted as the 
energy and the angular momentum, respectively, and they are constants of 
motion. 

\section{Quantum elementary systems of $\ONH$}\label{sec:qes}
We have combined both the theory of Mackey of induced representations 
\cite{Mac78} and the theory of Kirillov for nilpotent groups \cite{Kir76} 
in order to obtain all the different irreducible representations of $\ONH$. 
We take advantage of the structure of semi--direct product of a nilpotent 
subgroup, including extensions, space translations and boosts, times a 
subgroup containing time translations and rotations. 

In this case we can associate in a natural way coadjoint orbits and 
irreducible representations using an analogous of Kirillov's 
\emph{orbit method} \cite{Kir91}. We will expose in  some detail the 
procedure of constructing representations associated to the so--called 
\emph{maximal} orbits (those with $f,m\neq 0$ and $f\neq\pm m\tau$) and 
summarize the techniques for the rest, giving explicit expressions for all 
the cases. 

\fbox{\ref{eq:orbnh21a}}
We choose, among the different decompositions of $\ONH$, the following 
semidirect product 
\be
\ONH(2+1)=N \odot K,
\ee
where $N$ is the nilpotent subgroup including central extensions, space 
translations and boosts, while $K$ is the direct product of the time 
translations subgroup and rotations subgroup ($K=T\otimes R$). 

Firstly, let us calculate a representation for the subgroup $N$ using 
Kirillov's theory. Coadjoint action is reduced in $N$ to 
\be\label{eq:accnilnh}
\begin{split}
f'&=f, \\ m'&=m, \\ 
\mbf{p}'&=\mbf{p}-\frac{f}{\tau^2}\mbf{a}^{\pi /2}-m\mbf{v}, \\[.3cm]
\mbf{k}'&=\mbf{k}-f\mbf{v}^{\pi /2}+m\mbf{a}. \end{split}
\ee
The coadjoint orbits lead to different types of irreducible representations 
of $N$. We are interested here in those with $f\neq 0,\,m\neq 0$ and 
$f\neq\pm m\tau$, in order to be able to associate the representations that 
we will obtain with maximal orbits of the group. 

There are two invariants ($f,m$) of the action \eqref{eq:accnilnh}, so the 
corresponding orbits are diffeomorphic to $\R^4$ and representations are 
labeled by such constants. Let us choose an arbitrary point 
$u=(f,m,\mbf{0},\mbf{0})$ on the orbit $O_{f,m}$, and a maximal subalgebra 
$\mcal{L}$ of $\mcal{N}$ subordinate to $u$, i.e., $\langle 
u|[B,C]\rangle=0,\;\;\forall\;B,C\in \mcal{L}$. There are several options, 
all of them leading to equivalent representations, from which we select the 
subalgebra spanned by the generators 
$F,M,P_1-\frac{1}{\tau}K_1,P_2+\frac{1}{\tau}K_2$, i.e., 
\be\label{eq:subalmaximnh21}
\mcal{L}=\langle F,M,P_1-\frac{1}{\tau}K_1,P_2+\frac{1}{\tau}K_2\rangle.
\ee

It is easy now to find a 1D representation $\Delta$ of the group $L$, 
obtained via exponentiation of $\mcal{L}$, by means of duality 
\be
\Delta(l)=e^{i<u|B>}=e^{i(\xi f+\mu m)},
\ee
where we use the parametrization 
\be
l=(\xi,\mu,\delta_1,\delta_2)\,=\, e^B\,=\,e^{\bigl(\xi F+\mu M
        +\delta_1(P_1-\frac{K_1}{\tau})
        +\delta_2(P_2+\frac{K_2}{\tau})\bigr)}.
\ee

The support space of the representation of $N$ is the set of functions 
defined on the homogeneous space $N/L\equiv X$, $\mcal{F}(X,\C)$. Each 
coset is characterized by a point $\mbf{y}=(y_1,y_2)\in \R^2$ in such a way 
that we define the normalized Borel section 
\be
\ba{rccl}\sigma: & X & \longrightarrow & N \\
                 & \mbf{y} & \mapsto & (0,0,\mbf{0},\mbf{y}), \ea 
\ee
with $\pi\circ\sigma=\text{id}_{N/L}$, the canonical projection $\pi$ being 
defined by
\be
\ba{rccl}\pi: & N & \longrightarrow & X \\
              & (\alpha,\theta,\mbf{a},\mbf{v}) & \mapsto & 
                        \mbf{v}-\frac{\mbf{a}^r}{\tau}, \ea 
\ee
where $\mbf{a}^r=(-a_1,a_2)$. This defines the action of $N$ on $X$ through 
\be
g\mbf{y}:=\pi\bigl(g\sigma(\mbf{y})\bigr)
         =\mbf{v}+\mbf{y}-\frac{\mbf{a}^r}{\tau}. 
\ee
To finish the first step it only remains to induce the representation from 
$L$ to $N$, which is done using the fundamental relation 
\be
[D_{m,f}(g)\varphi](g\mbf{y})=\Delta (\sigma^{-1}(g\mbf{y})g
\sigma(\mbf{y}))\varphi(\mbf{y}), \qquad \forall g \in N.
\ee
This can be expressed in a suitable form as
\be\label{eq:repnilnh21a}
[D_{m,f}(g)\varphi](\mbf{y})
        =e^{if[\alpha+\frac{1}{2}(\mbf{v}\times\mbf{y})
                +\frac{1}{2\tau}(\mbf{v}-\mbf{y})\times\mbf{a}^r]}\;
        e^{im[\theta-\mbf{y}\mbf{a}-\frac{1}{2\tau}\mbf{a}^r\mbf{a}]}\;
        \varphi\Bigl(\mbf{y}-\mbf{v}+\frac{\mbf{a}^r}{\tau}\Bigr). 
\ee

The second step consists of the induction, from the preceding 
representation, of one for the whole group. For that purpose we must study 
the action of the subgroup $K=T\otimes R$ on the space of equivalence 
classes of representations of $N$, $\hat{N}$, which is parametrized by the 
values of $f$ and $m$. This action is carried out by conjugation, i.e., 
\be
D_{m,f}'(g):= D_{m,f}(k^{-1}gk),\qquad g\in 
                                N,\;k\in K. 
\ee
The group $K$ acts trivially on $\hat{N}$ as it is easy to see, and 
therefore, the little group $L_{m,f}$ coincides with $K$. Consequently, the 
representations of $N\odot L_{m,f}$, from which we achieve the induction, 
are truly representations of $\ONH$. More explicitly, we use the fact that 
$L_{m,f}$ is a direct product  to accomplish induction first with time 
translations and then with rotations. To start with, we construct a 
representation $D^t$ of $N\odot T$ by
\be
D^t(g):=D(t^{-1}gt),\quad g\in N,\; t\in T,
\ee
giving raise to 
\begin{multline}\label{eq:inductnh21}
[D^{t}_{m,f}(g)\varphi](\mbf{y})=\exp\{if[\alpha+\frac{1}{2}(\mbf{v}
        \cos\frac{b}{\tau}-\frac{\mbf{a}}{\tau}\sin\frac{b}{\tau})
        \times\mbf{y}\\[.3cm]
       +\frac{1}{2\tau}((\mbf{v}\cos\frac{b}{\tau}
        -\frac{\mbf{a}}{\tau}\sin\frac{b}{\tau})
        -\mbf{y})\times(\mbf{a}\cos\frac{b}{\tau}
        +\mbf{v}\tau\sin\frac{b}{\tau})^r]\}\\[.3cm] 
\times \exp\{im[\theta+\frac{1}{2}(\tau\mbf{v}^2
        -\frac{\mbf{a}^2}{\tau})\sin\frac{b}{\tau}\cos\frac{b}{\tau}
        -\mbf{a}\mbf{v}\sin^2\frac{b}{\tau}
        -\mbf{y}(\mbf{a}\cos\frac{b}{\tau}
        +\mbf{v}\tau\sin\frac{b}{\tau})\\[.3cm]
       -\frac{1}{2\tau}(\mbf{a}\cos\frac{b}{\tau}
        +\mbf{v}\tau\sin\frac{b}{\tau})^r(\mbf{a}\cos\frac{b}{\tau}
        +\mbf{v}\tau\sin\frac{b}{\tau})]\}\\[.3cm] 
\times\varphi\Bigl(\mbf{y}-\mbf{v}\cos\frac{b}{\tau}+\frac{\mbf{a}}{\tau}
        \sin\frac{b}{\tau}+\frac{1}{\tau}\bigl(\mbf{a}\cos\frac{b}{\tau}
        +\mbf{v}\tau\sin\frac{b}{\tau}\bigr)^r\Bigr). 
\end{multline}

The following step is to check the existence of an operator $W(b)$ 
realizing the equivalence in the sense of Mackey's theory, i.e., verifying 
\be\label{eq:opequiv}
D^t_{m,f}(g)=W^{-1}(b)\;D_{m,f}(g)\;W(b),\qquad 
        \forall g\in N,\;\forall t\in T.
\ee
This operator is reduced in most cases to exponentials of Casimir 
operators, which formally coincides with the invariants of the coadjoint 
action for the considered group ($N\odot T$). For the whole group $\ONH$ 
such invariants are 
\be\label{eq:invariants} \begin{split}
C_1&=\mbf{p}^2+\frac{\mbf{k}^2}{\tau^2}-2mh                               
                        +\frac{2}{\tau^2}f j,\\[.3cm] 
C_2&=\mbf{p}\times\mbf{k}+mj-f h. \end{split} 
\ee
Omitting rotations, this two invariants are reduced to one in order  to 
maintain the even dimension of the orbit and this new one can be obtained 
eliminating $j$ in the preceding equations. It turns out to be 
$C=C_1-\frac{2f}{\tau^2m}C_2$, i.e., 
\be 
C=\mbf{p}^2+\frac{\mbf{k}^2}{\tau^2}+\Bigl(\frac{2f^2}{\tau^2m}-2m\Bigr)h 
        -\frac{2f}{\tau^2m}(\mbf{p}\times\mbf{k}). 
\ee
If we solve for $h$ in this last expression and reinterpret it in terms of 
the algebra generators we get 
\be 
\hat{H}=\frac{\tau^2m}{2(\tau^2m^2-f^2)}\;\Bigl(\hat{\mbf{P}}^2
        +\frac{\hat{\mbf{K}}^2}{\tau^2}-\frac{2f}{\tau^2m}(\hat{\mbf{P}}
        \times\hat{\mbf{K}})-C\hat{I}\Bigr), 
\ee
where we have consider $\hat{M}=m$ and $\hat{F}=f$ since we are in the 
orbit $O_{f,m}$, while $\hat{\mbf{P}}$ and $\hat{\mbf{K}}$ can be 
calculated from 
\eqref{eq:repnilnh21a} and whose differential expressions are
\be\begin{split}
\hat{\mbf{P}}&=\frac{f}{2\tau}(\mbf{y}^{\frac{-\pi}{2}})^r-m\mbf{y}
        -\frac{i}{\tau}\nabla^{R}_{\mbf{y}},\\[.3cm]
\hat{\mbf{K}}&=\frac{f}{2}\mbf{y}^{\frac{-\pi}{2}}+i\nabla_{\mbf{y}}.
\end{split}
\ee

The intertwining operator verifying identity \eqref{eq:opequiv} can be 
defined by $W(b)=e^{i\hat{H}b}$. Thus, we get the induced representation 
$U_{m,f,C}$ for $N \odot T$ given by 
\be\begin{split}
[U_{m,f,C}(g)\varphi](\mbf{y})
   &=e^{if[\alpha+\frac{1}{2}(\mbf{v}\times\mbf{y})+\frac{1}{2\tau}
        (\mbf{v}-\mbf{y})\times\mbf{a}^r]}\;
        e^{im[\theta-\mbf{y}\mbf{a}
        -\frac{1}{2\tau}\mbf{a}^r\mbf{a}]}\\[.3cm]
   &\hspace{1cm}\times e^{i\frac{\tau^2m}{2(\tau^2m^2-f^2)}(\hat{\mbf{P}}^2
        +\frac{\hat{\mbf{K}}^2}{\tau^2}-\frac{2f}{\tau^2m}(\hat{\mbf{P}}
        \times\hat{\mbf{K}})-C)b}\;
        \varphi\Bigl(\mbf{y}-\mbf{v}+\frac{\mbf{a}^r}{\tau}\Bigr) \end{split}
\ee
where we have chosen a 1D irreducible unitary representation (i.u.r.) of 
$T$ ($b\in T\rightarrow e^{ibn},\;n\in\Z$) which is included in constant 
$C$. 

Following the same method we can complete the induction process and 
construct the i.u.r. for the whole group $\ONH$ departing from the ones 
just calculated. Since  $N\odot T$ is not abelian, we conjugate a generic 
element $g\in N\odot T$ with a rotation $R_{\phi} 
\in R$, obtaining 
\be
g^{\phi}=R^{-1}(\phi)gR(\phi) 
        =(\alpha,\theta,b,\mbf{a}^{-\phi},\mbf{v}^{-\phi},0). 
\ee
From here we construct the following representation of $N\odot T$ 
\be \begin{split}
[U_{m,f\,C}^{\phi}(g)\varphi](\mbf{y})
        &=e^{if[\alpha+\frac{1}{2}(\mbf{v}^{-\phi}
        \times\mbf{y})+\frac{1}{2\tau}(\mbf{v}^{-\phi}-\mbf{y})
        \times(\mbf{a}^{-\phi})^r]}\;
        e^{im[\theta-\mbf{y}\mbf{a}^{-\phi}-\frac{1}{2\tau}
        (\mbf{a}^{-\phi})^r\mbf{a}^{-\phi}]}\\[.3cm]
  &\hspace{1cm}\times e^{i\frac{\tau^2m}{2(\tau^2m^2-f^2)}[\hat{\mbf{P}}^2
        +\frac{\hat{\mbf{K}}^2}{\tau^2}-\frac{2f}{\tau^2m}(\hat{\mbf{P}}
        \times\hat{\mbf{K}})-C]b}\\[.3cm]
  &\hspace{1cm}\times \varphi\Bigl(\mbf{y}-\mbf{v}^{-\phi}
        +\frac{(\mbf{a}^{-\phi})^r}{\tau}\Bigr).\end{split}
\ee

Now, since rotations acts trivially on time translations (remember that 
$K=T\otimes R$), we can neglect the latter and consider $\tilde{g}\in 
NH/T$, with a considerable saving of calculation. We must find again an 
intertwining operator $W(\phi)$ such that 
\be\label{eq:operadorrotac}
U^{\phi}(\tilde{g})=W^{-1}(\phi) \;U(\tilde{g})\; W(\phi). 
\ee
We take $W(\phi)=e^{i\phi\hat{J}}$, where $\hat{J}$ is obtained on solving 
for $j$ in the expression \eqref{eq:invariants} of the invariant $C_2$, 
i.e., 
\be
\hat{J}=\frac{1}{m}(C_2-\hat{\mbf{P}}\times\hat{\mbf{K}}+f\hat{H}),
\qquad W(\phi)=e^{\frac{-i}{m}(\hat{\mbf{P}}\times\hat{\mbf{K}}-f\hat{H}
-C_2)\phi}.
\ee

Finally, the expression for the representation associated to the orbit 
labeled by $O_{f,m}^{C_1,C_2}$ turns out to be
\be\label{eq:repnh21a}
\begin{split} [U^{C_1,C_2}_{m,f}(g)\varphi](\mbf{y}) 
   &=e^{if[\alpha+\frac{1}{2}(\mbf{v}\times\mbf{y})
        +\frac{1}{2\tau}(\mbf{v}-\mbf{y})\times\mbf{a}^r]}\;
        e^{im[\theta-\mbf{y}\mbf{a}-\frac{1}{2\tau}\mbf{a}^r\mbf{a}]}\\[.3cm]
   &\hspace{1cm}\times e^{i\frac{\tau^2m}{2(\tau^2m^2-f^2)}[\hat{\mbf{P}}^2
        +\frac{\hat{\mbf{K}}^2}{\tau^2}-\frac{2f}{\tau^2m}(\hat{\mbf{P}}
        \times\hat{\mbf{K}})-C_1+\frac{2f}{\tau^2m}C_2]b}\\[.3cm]
   &\hspace{1cm}\times e^{\frac{-i}{m}(\hat{\mbf{P}}
        \times\hat{\mbf{K}}-f\hat{H}-C_2)\phi}\;
        \varphi\Bigl(\mbf{y}-\mbf{v}+\frac{\mbf{a}^r}{\tau}\Bigr), \end{split}
\ee
where the constants $C_2$ and $C_1$ in the exponentials include, 
respectively, an integer number $n'$ corresponding to the 1D i.u.r. 
$e^{in'\phi}$ of $SO(2)$ ($\phi\rightarrow e^{in'\phi},\;n'\in\Z$) and a 
real number associated to the i.u.r. of $T$ before mentioned. 

\fbox{\ref{eq:orbnh21b}}
When we consider $f=m\tau$, the coadjoint orbits for the nilpotent factor 
$N$ are characterized by the invariants $f$ and $\mbf{x}=\mbf{p}+ 
\frac{\mbf{k}^{\pi/2}}{\tau}$, so every i.u.r. of $N$ is labeled by 
the values $(f,\mbf{x})$. In this case, the maximal subalgebra subordinate 
to a point on the orbit must be 4D, for example, we take 
\be
\mcal{L}=\langle F,\;P_1-\frac{1}{\tau}K_1,\;P_2+\frac{1}{\tau}K_2,\;P_1
        -\frac{1}{\tau}K_2\rangle.
\ee
This leads to the following representation for $N$  
\be\label{eq:repfx}\begin{split}
[D_{f,\mbf{x}}(n)\varphi](y)
&=e^{if[\alpha+y(v_1-\frac{a_2}{\tau})-\frac{1}{2}(v_1^2+v_1v_2
        +\frac{a_2^2-a_1a_2}{\tau^2})+\frac{v_1}{\tau}(a_2-a_1)]}\\[.3cm]
&\qquad\times e^{i\mbf{x}\mbf{a}}\;\varphi\bigl(y-v_1-v_2-\frac{a_1}{\tau}
        +\frac{a_2}{\tau}\bigr),\end{split}
\ee
defined on $\mcal{L}^2(\R)$.

The action of $K$ on $\hat{N}$ is given by 
$(f',\mbf{x}')=(f,\mbf{x}^{(\frac{b}{\tau}+\phi)})$, so there are two kinds 
of strata depending on the value of the invariant $\mbf{x}^2$. In this case 
we choose $\mbf{x}^2\neq 0$ in order to associate this value with that of 
constant $C_3$ defining the strata of orbits we are dealing with (the case 
$C_3=0$ will be studied later on  for orbits 
\eqref{eq:orbnh21d}). 

The little group of the orbits of the strata characterized by 
$\mbf{x}^2\neq 0$ is $L_-=\{(b,\phi)\in T\otimes 
R\;|\;\frac{b}{\tau}+\phi=0\}$, which is generated by $H-\frac{1}{\tau}J$. 
The operator $W(l_-)$ carrying out the equivalence between representation 
$D_{f,\mbf{x}}$ and its conjugate by the action of $K$, $D_{f,\mbf{x}}^k$, 
is 
\be\label{eq:uvwdoble}
W\bigl(b,\frac{-b}{\tau}\bigr)=e^{ib\widehat{(H-\frac{1}{\tau}J)}}=e^{ib\frac{\tau}{4f}
        (\hat{\mbf{P}}^2+\frac{\hat{\mbf{K}}^2}{\tau^2}
        +\frac{2}{\tau}(\hat{\mbf{P}}\times\hat{\mbf{K}})-C_4\hat{I})},
\ee 
where the differential representation of $\hat{\mbf{P}}$ and 
$\hat{\mbf{K}}$ are obtained from  
\eqref{eq:repfx}. Thus, the representation $\Upsilon$ of $N\odot L_-$ from which 
to induce is the tensor product of the representation $e^{ibC}$ of $L_-$ 
times $D_{f,\mbf{x}}(n)W(l_-)$ of $N$. 

The homogeneous space $\ONH/(N\odot L_-)$ is isomorphic to $\R$ and its 
elements will be denoted by $t$, with canonical projection 
$\pi:\ONH\longrightarrow \ONH/(N\odot L_-)$ given by 
$\pi(\alpha,b,\mbf{a},\mbf{v},\phi)=\frac{b}{\tau}+\phi$ and normalized 
Borel section defined by $s(t)=(0,\tau t,\mbf{0},\mbf{0},0)$. 

The i.u.r. of the group $\ONH$ induced by the representation $\Upsilon$ of 
$N\odot L_-$ is expressed by 
\be\label{eq:repnh21b}
\begin{split}
[U_f^{C_3,C_4}\Psi](t)=\Upsilon(\eta)\;\Psi(t-\frac{b}{\tau}-\phi)
\end{split}
\ee
where $\Psi\in \mcal{L}^2(\R)$, and $\eta=s^{-1}(t)gs(g^{-1}t)$, explicitly 
\be\begin{split}
\eta&=\Bigl(\alpha+\frac{1}{2}\cos(t-\frac{b}{\tau}-\phi)\sin(t-\frac{b}{\tau}-\phi)
        (\mbf{v}^2-\frac{\mbf{a}^2}{\tau^2})-\frac{\mbf{a}\mbf{v}}{\tau}
        \sin^2(t-\frac{b}{\tau}-\phi), \\[.3cm]
&\phantom{=(bcd} -\tau\phi,\;\mbf{a}\cos(t-\frac{b}{\tau}-\phi)+\tau\mbf{v}
        \sin(t-\frac{b}{\tau}-\phi),\; \\[.3cm]
&\phantom{=(bcde} \mbf{v}\cos(t-\frac{b}{\tau}-\phi)
        -\frac{\mbf{a}}{\tau}\sin(t-\frac{b}{\tau}-\phi),\; 
        \phi\Bigr). \end{split}
\ee

\fbox{\ref{eq:orbnh21c}}
This case is analogous to the preceding one. Invariants characterizing the 
orbits for the nilpotent part are $(f,\,\mbf{y}=\mbf{p}- 
\frac{\mbf{k}^{\pi/2}}{\tau})$ and we can choose the subalgebra
\be
\mcal{L}=\langle F,\;P_1-\frac{1}{\tau}K_1,\;P_2+\frac{1}{\tau}K_2,\;P_1
        +\frac{1}{\tau}K_2\rangle
\ee
in order to obtain the following i.u.r. for $N$ defined on $\mcal{L}^2(\R)$ 
\be\label{eq:repfy}\begin{split}
[D_{f,\mbf{y}}(n)\varphi](x)
&=e^{if[\alpha-x(v_1+\frac{a_2}{\tau})+\frac{1}{2}(v_1^2-v_1v_2
        +\frac{a_2^2+a_1a_2}{\tau^2})+\frac{v_1}{\tau}(a_1+a_2)]}\\[.3cm]
&\qquad\times e^{i\mbf{y}\mbf{a}}\;\varphi(x-v_1+v_2-\frac{1}{\tau}
        (a_1+a_2)).\end{split}
\ee
The action of $K$ on $\hat{N}$ is $(f',\mbf{y}') 
=(f,\mbf{y}^{(\frac{b}{\tau}-\phi)})$. The little group of the orbit 
characterized by $\mbf{y}^2=C_3'\neq 0$ is generated by 
$H+\frac{1}{\tau}J$, i.e., 
\be
L_+=\{(b,\phi)\in T\otimes R\;|\;\frac{b}{\tau}-\phi=0\}.
\ee

The operator $W(l_+)$ relating representations $D_{f,\mbf{y}}$ and 
$D^k_{f,\mbf{y}},\;k\in K$ is 
\be
W(b,\frac{b}{\tau})=e^{ib\widehat{(H+\frac{1}{\tau}J)}}=e^{-ib\frac{\tau}{4f}
        (\hat{\mbf{P}}^2+\frac{\hat{\mbf{K}}^2}{\tau^2}
        -\frac{2}{\tau}(\hat{\mbf{P}}\times\hat{\mbf{K}})-C_4'\hat{I})},
\ee 
where the values of $\hat{\mbf{P}}$ and $\hat{\mbf{K}}$ are implicit in 
\eqref{eq:repfy}. The representation $\Upsilon$ of $N\odot L_+$ from which 
we induce is the tensor product of the representation of $L_+$, i.e., 
($e^{ibC'}$) times $D_{f,\mbf{y}}(n)W(l_+)$. 

Finally, the induced representation is 
\be\label{eq:repnh21c}
[U_f^{C_3',C_4'}\Psi](t)=\Upsilon(\eta')\;
        \Psi\bigl(t-\frac{b}{\tau}+\phi\bigr)
\ee
where $\Psi\in\mcal{L}^2(\R)$ and the element of induction, $\eta$, is 
\be\begin{split}
\eta'
&=\Bigl(\alpha+\frac{1}{2}\cos(t-\frac{b}{\tau}+\phi)\sin(t-\frac{b}{\tau}+\phi)
        (\mbf{v}^2-\frac{\mbf{a}^2}{\tau^2})-\frac{\mbf{a}\mbf{v}}{\tau}
        \sin^2(t-\frac{b}{\tau}+\phi), \\[.3cm]
&\hspace{1.5cm} \tau\phi,\;\mbf{a}\cos(t-\frac{b}{\tau}+\phi)+\tau\mbf{v}
        \sin(t-\frac{b}{\tau}+\phi),\; \\[.3cm]
&\hspace{1.5cm} \mbf{v}\cos(t-\frac{b}{\tau}+\phi)
        -\frac{\mbf{a}}{\tau}\sin(t-\frac{b}{\tau}+\phi),\; 
        \phi\Bigr). \end{split}
\ee

\fbox{\ref{eq:orbnh21d}}
This case completes the study started in \ref{eq:orbnh21b} since 
corresponds to take $C_3\equiv\mbf{x}^2=0$. The action of $K$ on $\hat{N}$ 
is trivial and the little group coincides with $K$. Therefore, the 
representations from which we realize the induction are true 
representations of $\ONH$ and it is not necessary to take the last step of 
case \ref{eq:orbnh21b}. Consequently, we must consider an i.u.r. of the 
abelian group $K$,  $\rho_{\kappa_1,\kappa_2}(b,\phi)= 
e^{i(b\kappa_1+\phi\kappa_2)}$, the representation of the nilpotent part 
($D_{f,\mbf{x}=\mbf{0}}$) and the intertwining operator $W(b,\phi)$ 
realizing the equivalence in the sense of Mackey. This operator can be 
written in terms of the generators $H+\frac{1}{\tau}J$ and 
$H-\frac{1}{\tau}J$, which can be directly obtained from the coadjoint 
action invariants $C_4$ and $C_5$, explicitly 
\be
W(b,\phi)=e^{\frac{i}{2}(b+\tau\phi)C_5}\;e^{i\frac{\tau}{8f}(b-\tau\phi)
        [\hat{\mbf{P}}^2+\frac{\hat{\mbf{K}}^2}{\tau^2}
        +\frac{2}{\tau}(\hat{\mbf{P}}\times\hat{\mbf{K}})-C_4\hat{I}]}.
\ee
The final expression for the i.u.r. is 
\be\label{eq:repnh21d}
U_f^{C_4,C_5}(g)=\rho_{\kappa_1,\kappa_2}(k)\otimes 
D_{f,\mbf{x}=\mbf{0}}(n)W(b,\phi). 
\ee

\fbox{\ref{eq:orbnh21e}}
This case can be solved as the preceding one and complete the case 
\ref{eq:orbnh21c} since now we consider $C_3'\equiv\mbf{y}^2=0$. The little 
group of the action of $K$ on $\hat{N}$ is again $T\otimes R$. Hence, the 
corresponding i.u.r. is 
\be\label{eq:repnh21e}
U_f^{C_4',C_5'}(g)=\rho_{\kappa_1,\kappa_2}(k)\otimes 
D_{f,\mbf{y}=\mbf{0}}(n) W(b,\phi), 
\ee
where $\rho$ is the same character of previous case and the operator $W$ is  
now 
\be
W(b,\phi)=e^{\frac{i}{2}(b-\tau\phi)C_5'}\;e^{i\frac{\tau}{8f}(b+\tau\phi)
        [\hat{\mbf{P}}^2+\frac{\hat{\mbf{K}}^2}{\tau^2}
        -\frac{2}{\tau}(\hat{\mbf{P}}\times\hat{\mbf{K}})-C_4'\hat{I}]}.
\ee 

\fbox{\ref{eq:orbnh21f}}
The construction of this representation follows the procedure used for 
maximal orbits. In this case, there is only one invariant, $m$, for the 
coadjoint action of the nilpotent factor of the group, so the coadjoint 
orbits are isomorphic to $\R^4$ and representations are labeled by this 
constant. 

On the other hand, condition $f=0$ is equivalent to the fact that 
generators of space translations commute, $\text{i.}\text{e.},\; 
[P_1,P_2]=0$. This allows us to choose the maximal subordinate subalgebra 
to a point in the orbit as $\mcal{L}=<M,P_1,P_2>$, which facilitates 
calculations and leads to a representation equivalent to the obtained by 
using subalgebra 
\eqref{eq:subalmaximnh21}, which would have an expression similar to that 
found in \eqref{eq:repnh21a}. 

The i.u.r. is defined on $\mcal{L}^2(\R^2)$, with velocity kind argument 
and takes the form 
\be\label{eq:repnh21f}
[U_{m}^{C_1,C_2}(g)\varphi](\mbf{y})
        =e^{im[\theta-\mbf{y}\mbf{a}]}\;e^{\frac{i}{2m}(\hat{\mbf{P}}^2
        +\frac{\hat{\mbf{K}}^2}{\tau^2}-C_1\hat{I})b}\;
        e^{\frac{-i}{m}(\hat{\mbf{P}}\times\hat{\mbf{K}}-C_2\hat{I})\phi}\;
        \varphi(\mbf{y}-\mbf{v}),
\ee
where constants $C_1,C_2$ include the parameters labeling the i.u.r. of $T$ 
and rotations. In section \ref{subsec:ejemplo} we use the differential form 
of operators $\hat{\mbf{P}},\hat{\mbf{K}}$ acting on functions in 
$\mcal{L}^2(\R^2)$ as $\hat{\mbf{P}}=-m\mbf{y}$, 
$\hat{\mbf{K}}=i\nabla_{\mbf{y}}$ in order to construct the SW 
correspondence. 
 
\fbox{\ref{eq:orbnh21g}}
The process is completely analogous to the preceding one interchanging the 
roles played by $m$ and $f$. There are only two differences, the first one 
is the choice of the subalgebra $\mcal{L}=\langle F, P_1-\frac{1}{\tau}K_1, 
P_2+\frac{1}{\tau}K_2\rangle$, the same as in the maximal case. The second 
one appears as a consequence of $m=0$ in the orbit invariants when defining 
the operators $\hat{H}$ and $\hat{J}$, although the option is obvious and 
can be noticed in the final result, which again is a representation acting 
on $\mcal{L}^2(\R^2)$ 
\be\label{eq:repnh21g}
\begin{split}[U_{f}^{C_1,C_2}(g)\varphi](\mbf{y})
   &=e^{if[\alpha+\frac{1}{2}(\mbf{v}\times\mbf{y})+\frac{1}{2\tau}
        (\mbf{v}-\mbf{y})\times\mbf{a}^r]}\;
     e^{i\frac{1}{f}(\hat{\mbf{P}}\times\hat{\mbf{K}}-C_2)b}\\[.3cm]
   &\hspace{1cm}\times e^{i\frac{\tau^2}{2f}(\hat{\mbf{P}}^2
        +\frac{\hat{\mbf{K}}^2}{\tau^2}-C_1)\phi}\;
        \varphi\Bigl(\mbf{y}-\mbf{v}+\frac{\mbf{a}^r}{\tau}\Bigr), \end{split}
\ee
where $C_1$ and $C_2$ include the labels of the i.u.r. of one--parameter 
groups as in the previous cases. 

\fbox{\ref{eq:orbnh21h}}
The remaining cases correspond to the unextended group, i.e., $f=m=0$. 

The i.u.r. of the abelian kernel $N$ (space translations and boosts) are 
labeled by the characters. These 1D representations are defined by 
\be
D_{\brho,\bvarkappa}(\mbf{a},\mbf{v}):=e^{i(\brho\mbf{a}+\bvarkappa\mbf{v})}, 
        \qquad \brho,\,\bvarkappa\,\in\R^2.
\ee
The action of the other group factor, $K$, on the set of characters 
$(\brho,\bvarkappa)$ is easily computed and results to be 
\be\begin{split}
\brho'&=\brho^{\phi}\cos\frac{b}{\tau}
        -\frac{1}{\tau}\bvarkappa^{\phi}\sin{b}{\tau},\\[.3cm]
\bvarkappa'&=\tau\brho^{\phi}\sin\frac{b}{\tau}
        +\bvarkappa^{\phi}\cos{b}{\tau}. \end{split}
\ee
This action is formally identical to the restriction to $K$ of the 
coadjoint action \eqref{eq:coadnh21} for $(\mbf{p},\mbf{k})$ and, 
therefore, its invariants are either identical. So we can write 
\be\begin{split}
\brho^2+\frac{\bvarkappa^2}{\tau^2}=I_1,\\[.3cm]
\brho\times\bvarkappa=I_2. \end{split}
\ee
Depending on the values assigned to $I_1$ and $I_2$, we have some of the 
possible four kinds of orbits and hence, of representations.

As first case, we suppose $I_2\neq\pm\frac{\tau}{2}I_1$, which results in  
2D character orbits, whose little group is trivial. The induced 
representations act on functions defined on the homogeneous space  
$NH/N\backsimeq \R^2$, where we take the section $s(\mbf{t})= 
(t_1,\mbf{0},\mbf{0},t_2)$ with projection $\pi(b,\mbf{a},\mbf{v},\phi)= 
(b,\phi)$. The action of the group $NH$ on this space is 
$g\mbf{t}=(t_1+b,t_2+\phi)$ and the induction element is 
\be\begin{split}
s^{-1}(\mbf{t})\,g\,s(g^{-1}\mbf{t})
&=(-t_1,\mbf{0},\mbf{0},-t_2)(b,\mbf{a},\mbf{v},\phi)
        (t_1-b,\mbf{0},\mbf{0},t_2-\phi)\\[.3cm]
&=(0,\,\mbf{a}^{-t_2}\cos\frac{t_1-b}{\tau}
        +\tau\mbf{v}^{-t_2}\sin\frac{t_1-b}{\tau},\\[.3cm]
&\qquad \mbf{v}^{-t_2}\cos\frac{t_1-b}{\tau}
        -\frac{1}{\tau}\mbf{a}^{-t_2}\sin\frac{t_1-b}{\tau},\,0).\end{split}
\ee
The induced i.u.r. is expressed as 
\be\label{eq:repnh21h}
\begin{split} [U_{C_1,C_2}(g)\varphi](\mbf{t}) 
&=e^{i\brho(\mbf{a}^{-t_2}\cos\frac{t_1-b}{\tau} 
        +\tau\mbf{v}^{-t_2}\sin\frac{t_1-b}{\tau})}\\[.3cm]
&\qquad\times e^{i\bvarkappa(\mbf{v}^{-t_2}\cos\frac{t_1-b}{\tau}
        -\frac{1}{\tau}\mbf{a}^{-t_2}\sin\frac{t_1-b}{\tau})}
        \;\varphi(t_1-b,t_2-\phi), \end{split}
\ee
with $C_1=\brho^2+\frac{\bvarkappa^2}{\tau^2}$ and 
$C_2=\brho\times\bvarkappa$. 

\fbox{\ref{eq:orbnh21i}}
Departing from the preceding case and taking  $I_2=(\tau/2)I_1$, we get the 
new invariant $\brho+\frac{\bvarkappa^{\pi/2}}{\tau}=\mbf{0}$ and conclude 
that the action on the characters is reduced to the rotation 
\be
\bvarkappa'=\bvarkappa^{(\frac{b}{\tau}-\phi)},
\ee
and hence, the corresponding little group, generated by 
$H+\frac{1}{\tau}J$, is 
\be
L_+=\{(b,\phi)\in T\otimes R\;|\;\frac{b}{\tau}-\phi=0\}.
\ee 
The i.u.r. of  $L_+$ are $\Delta_C(l)=e^{ibC}$, where $b$ is the parameter 
corresponding to the element of $L_+$ written as $(b,\,b/\tau)$. The real 
constant $C$ can be associated to the value of constant $C_5$ of the 
coadjoint orbit giving the value of the invariant $h+j/\tau$. 

Therefore, we induce from the representation of $N\odot L_+$, 
$U(n,h_+)=e^{ibC_5}\; e^{i(\brho\mbf{a}+\bvarkappa\mbf{v})}$. The 
homogeneous space $NH/(N\odot L_+)$ is now isomorphic to $\R$, and we can 
define the section $s(t)=(\tau t,\mbf{0},\mbf{0},0)$ with canonical 
projection $\pi(b,\mbf{a},\mbf{v},\phi)=\frac{b}{\tau}-\phi$. The induction 
element is
\be\begin{split}
s^{-1}(t)\,g\,s(g^{-1}t)
&=(\tau\phi,\,\mbf{a}\cos(t-\frac{b}{\tau}+\phi)
        +\tau\mbf{v}\sin(t-\frac{b}{\tau}+\phi),\\[.3cm]
&\qquad \mbf{v}\cos(t-\frac{b}{\tau}+\phi)
        -\frac{1}{\tau}\mbf{a}\sin(t-\frac{b}{\tau}+\phi),\,\phi),\end{split}
\ee
and the representation is
\be\label{eq:repnh21i}
\begin{split} [U_{C_1,C_5}(g)\varphi](t) &=e^{i\tau\phi 
C_5}\;e^{i\brho[\mbf{a}\cos(t-\frac{b}{\tau}+\phi) 
        +\tau\mbf{v}\sin(t-\frac{b}{\tau}+\phi)]}\\[.3cm]
&\qquad\times e^{i\bvarkappa[\mbf{v}\cos(t-\frac{b}{\tau}+\phi)
        -\frac{1}{\tau}\mbf{a}\sin(t-\frac{b}{\tau}+\phi)]}
        \;\varphi\bigl(t-\frac{b}{\tau}+\phi\bigr), \end{split}
\ee
with $C_1=\brho^2+\frac{\bvarkappa^2}{\tau^2}$ and 
$\brho+\frac{\bvarkappa^{\pi/2}}{\tau}=\mbf{0}$.

\fbox{\ref{eq:orbnh21j}}
Construction process is identical to the previous case, so we do not repeat 
it here. We only have to notice that the little group is now generated by  
$H-\frac{1}{\tau}J$, leading to the i.u.r. 
\be\label{eq:repnh21j} 
\begin{split} [U_{C_1,C_5'}(g)\varphi](t) &=e^{-i\tau\phi 
C_5'}\;e^{i\brho[\mbf{a}\cos(t-\frac{b}{\tau}-\phi) 
        +\tau\mbf{v}\sin(t-\frac{b}{\tau}-\phi)]}\\[.3cm]
&\qquad\times e^{i\bvarkappa[\mbf{v}\cos(t-\frac{b}{\tau}-\phi)
        -\frac{1}{\tau}\mbf{a}\sin(t-\frac{b}{\tau}-\phi)]}
        \;\varphi\bigl(t-\frac{b}{\tau}-\phi\bigr), \end{split}
\ee
with $C_1=\brho^2+\frac{\bvarkappa^2}{\tau^2}$ and 
$\brho-\frac{\bvarkappa^{\pi/2}}{\tau}=\mbf{0}$.

\fbox{\ref{eq:orbnh21k}}
This last case corresponds to the identity character 
($\brho=\bvarkappa=0$). So, the little group is $T\otimes R$ and, 
therefore, the i.u.r. is reduced to the character of this abelian group, 
i.e., 
\be\label{eq:repnh21k}
U_{h,j}(g)=e^{ibh}\;e^{i\phi j}.
\ee

\section{An example of Moyal Quantization}\label{sec:quant}
Quantization of elementary systems has been a subject of interest in 
Physics as well as in Mathematics during last decades. It is due to the 
fact that it seeks to answer the open questions about the deep meaning of 
quantum theory and its relation with the classical one, as well as it tries 
to find  a mathematical structure able to globally formalize quantum 
phenomena. We can mention different approaches to the problem, such as 
geometric quantization \cite{Sni80,Woo92}, quantization on groups 
\cite{AA82,AA87}, quantization by deformation \cite{BFF78} or Fedosov's 
quantization \cite{Fed94}. All of them are conditioned by the formalism 
that describe quantum systems. In this sense, we consider here the 
so--called Moyal quantization, which is based on the Moyal formulation of 
Quantum Mechanics or phase space formalism \cite{Moy49}, whose main 
ingredient is the consideration of both observables and states as 
(generalized) functions on a given phase space, introducing quantum 
concepts by means of an associative but non commutative product called
\co twisted product''. 

In particular, we use the Stratonovich--Weyl (SW) correspondence, which 
departing from the ideas of Stratonovich \cite{Str57} has been developed 
during last years for different kinds of systems 
\cite{VG89,CGV90,GMN91,MO96,AMO96}. We refer the reader to the last three 
references for more details or to \cite{Gad95,AMO98} for a review paper. 

As an example of how this quantization scheme works for the classical 
elementary systems that we described in section \ref{sec:orb}, we 
accomplish in next subsection the quantization of orbits of kind 
\ref{eq:orbnh21f}. 

\subsection{Example of quantization}\label{subsec:ejemplo}
Let us consider one orbit of type \ref{eq:orbnh21f} labeled by the three 
invariants $m,C_1$ and $C_2$, $O_{C_1,C_2}^m$. We already saw that 
$(\mbf{q}=\mbf{k}/m,\:\mbf{p})$ constitute a set of canonical coordinates 
($\{q_i,p_j\}=\delta_{ij}$, $\{q_1,q_2\}=\{p_1,p_2\}=0$). Moreover, we can 
evaluate $h$ and $j$ departing from the values of 
$\mbf{q},\;\mbf{p},\;m,\;C_1$ and $C_2$ obtaining 
\be  
h=\frac{\mbf{p}^2}{2m}+\frac{m\mbf{q}^2}{2\tau^2}-\frac{C_1}{2m},\qquad
j=\frac{C_2}{m}-\mbf{p}\times\mbf{q}.
\ee
Thus, the orbit is isomorphic to $\R^4$.

We can fix as origin in the orbit the point $(\mbf{q} 
=\mbf{0},\;\mbf{p}=\mbf{0})$ and define on it the autoadjoint parity operator   
\be
[\Om(\mbf{0},\mbf{0})\varphi](\mbf{y}):=2^2\,\varphi(-\mbf{y})
\ee
(see Ref. \cite{GMN91} for more details). This definition of 
$\Om(\mbf{0},\mbf{0})$ is an Ansatz for the SW kernel that we must extend 
to the rest of the orbit using covariance property (this kernel is the 
central object of the SW correspondence, that relates functions on the 
considered phase space with operators on a suitable Hilbert space), i.e., 
\be
\Om(\mbf{q},\mbf{p})\equiv\Om\big(g(\mbf{0},\mbf{0})\big)
        =U(g)\Om(\mbf{0},\mbf{0})U(g^{-1}),
\ee
where $g(\mbf{0},\mbf{0})=(\mbf{q},\mbf{p})$. One such element $g$ of 
$\ONH$ is 
\be
g_{\mbf{qp}}=(0,0,\mbf{q},-\mbf{p}/m,0).
\ee
We get the SW kernel at any point $(\mbf{q},\mbf{p})$ as
\be 
[\Om(\mbf{q},\mbf{p})\varphi](\mbf{y}) 
        =2^2\;e^{-2i\mbf{q}(m\mbf{y}+\mbf{p})}\; 
        \varphi(-\mbf{y}-2\frac{\mbf{p}}{m}).
\ee
This definition of $\Om(\mbf{q},\mbf{p})$ is independent of the choice of 
the group element $g$ that moves $(\mbf{0},\mbf{0})$ to 
$(\mbf{q},\mbf{p})$, in other words, $\Om$ verifies the covariance property
as we can see in the following way: the isotopy group of 
$(\mbf{0},\mbf{0})$ is 
\be 
\Gamma_{(0,0)}\equiv\{g\in G\;|\;g=(\theta,b,\mbf{0},\mbf{0},\phi)\},
\ee
and the restriction of the representation $U_m^{C_1,C_2}$ of $\ONH$ 
\eqref{eq:repnh21f} to the isotopy group commutes with $\Om(\mbf{0},\mbf{0})$
as it is easy to check. It was proved in Ref. \cite{GMN91} that this 
commutation is a necessary and sufficient condition to assure the 
covariance property. 

The kernel $\Om$ also verifies the traciality property, which is checked as 
follows 
\be \begin{split} 
\TR[\Om(\mbf{0},\mbf{0})\Om(\mbf{q},\mbf{p})] 
&=\int_{\R^2}\langle\mbf{y}|\Om(\mbf{0},\mbf{0})\Om(\mbf{q},\mbf{p})
        |\mbf{y}\rangle\;d\mbf{y}\\[.3cm]
&=2^4\int_{\R^2}e^{2i\mbf{q}(\mbf{p}-m\mbf{y})}
        \langle\mbf{y}|\mbf{y}-\frac{2}{m}\mbf{p}\rangle\;d\mbf{y}\\[.3cm]
&=2^2m^2\delta(\mbf{p})\int_{\R^2}e^{-2im\mbf{q}\mbf{y}}\;d\mbf{y}
        =\delta(\mbf{q})\;\delta(\mbf{p}).\end{split}
\ee
This property is crucial in the theory because it allows to get an 
inversion formula for the SW correspondence and to evaluate quantum mean 
values by integration of corresponding functions along the phase space.

It is also easy to find an invariant measure $d\mu(\mbf{u})$ on the orbit 
with expression $d\mu(\mbf{u})=d\mbf{q}\,d\mbf{p}$ that we use to integrate 
on the orbit (phase space).

Finally, we conclude quantization process calculating the tri--kernel of 
the SW correspondence, that turns out to be 
\be 
\TR[\Om(\mbf{q},\mbf{p})\Om(\mbf{q}',\mbf{p}')\Om(\mbf{q}'',\mbf{p}'')]
        =2^4\;e^{2i\mbf{q}(\mbf{p}'-\mbf{p}'')}\;
        e^{2i\mbf{q}'(\mbf{p}''-\mbf{p})}\;
        e^{2i\mbf{q}''(\mbf{p}-\mbf{p}')},
\ee
and provide us with the integral kernel which defines the twisted product.

We must notice here that this scheme of quantization can be accomplished 
for all kind of orbits associated to $NH(2+1)$ groups, although due to the 
geometry of some orbits it is necessary to introduce additional machinery 
related with quantization of cylindric orbits that is out of the scope of 
this article and we postpone it to subsequent papers.  

\section{Concluding remarks}\label{sec:remarks}
We briefly discuss here the significance of the second extension associated 
with parameter $f$ that appeared in section \ref{sec:orb}. This extension 
does not exists for the groups $NH$ in (1+1) and (3+1) dimensions. In fact, 
a similar situation appears for (2+1) Galilei group in \cite{BGO92}. As we 
pointed out in the introduction, this difference is easily understood 
taking into account that Newton--Hooke groups $NH_{\pm}$ are 
space--velocity contractions of the de Sitter groups $dS_{\pm}$, i.e., $NH$ 
groups are to the  $dS$ ones as Galilei group is to  Poincar' one 
\cite{BL68,DD72}. 

In table \ref{tab:conm21} we present the commutation relations for the Lie 
algebras of the four groups above mentioned (in the second part of the 
table we show the central extension structure). From them we see that 
contraction $dS_{\pm}\longrightarrow NH_{\pm}$ is accomplished in the limit 
$c\rightarrow\infty,\;R\rightarrow\infty$ while the quotient $c/R$ is 
maintained constant ($c$ can be interpreted as the light speed and $R$ as 
the curvature radius of the de Sitter universe).  
\begin{table}[htb] 
\begin{center}
\setlength{\extrarowheight}{8pt}
\begin{tabular}{|l|l|l|l|}\hline
        de Sitter & Newton--Hooke & Galilei & Poincar' \\ \hhline{|=|=|=|=|}
        $[J,K_i]=\epsilon_{ij}K_j$ & $[J,K_i]=\epsilon_{ij}K_j$ & 
        $[J,K_i]=\epsilon_{ij}K_j$ & $[J,K_i]=\epsilon_{ij}K_j$ \\ 
        $[J,P_i]=\epsilon_{ij}P_j$ & $[J,P_i]=\epsilon_{ij}P_j$ & 
        $[J,P_i]=\epsilon_{ij}P_j$ & $[J,P_i]=\epsilon_{ij}P_j$ \\ 
        $[J,H]=0$ & $[J,H]=0$ & $[J,H]=0$ & $[J,H]=0$ \\
        $[K_i,K_j]=-\epsilon_{ij}\frac{1}{c^2}J$ & $\mbf{[K_i,K_j]=0}$ 
                & $\mbf{[K_i,K_j]=0}$ & $[K_i,K_j]=-\epsilon_{ij}J$ \\
        $[K_i,P_j]=\delta_{ij}\frac{1}{c^2}H$ & $\mbf{[K_i,P_j]=0}$ & 
        $\mbf{[K_i,P_j]=0}$
                & $[K_i,P_j]=\delta_{ij}H$ \\
        $[K_i,H]=P_i$ & $[K_i,H]=P_i$ & $[K_i,H]=P_i$ & $[K_i,H]=P_i$ \\
        $[P_i,P_j]=\mp\epsilon_{ij}\frac{1}{R^2}J$ & $\mbf{[P_i,P_j]=0}$ 
                & $[P_i,P_j]=0$ & $[P_i,P_j]=0$ \\
        $[P_i,H]=\mp(\frac{c}{R})^2K_i$ & $[P_i,H]=\mp\frac{1}{\tau^2}K_i$ 
                & $[P_i,H]=0$ & $[P_i,H]=0$ \\[.3cm] \hhline{|=|=|=|=|}
        \multirow{4}{30mm}%
        {\parbox{27mm}{\hspace{0pt}\noindent trivial \phantom{exten}extensions}} 
        & $[K_i,P_j]=\delta_{ij}M$ & $[K_i,P_j]=\delta_{ij}M$ & \\
        & $[K_i,K_j]=\epsilon_{ij}F$ & $[K_i,K_j]=\epsilon_{ij}F$ &
        \raisebox{-1ex}[0cm][0cm]%
          {\parbox{27mm}{\hspace{0pt}\noindent trivial \phantom{exten}extensions}} \\
        & $[P_i,P_j]=\epsilon_{ij}\frac{1}{\tau^2}F$ & & \\
        & $([H,J]=L)$ & $([H,J]=L)$ & \\[.3cm] \hline
\end{tabular}
\caption{Commutation relations in (2+1) dimensions.}\label{tab:conm21}
\end{center}
\end{table} 
In this sense, the interpretation of extension $f$ is parallel to that made 
for  Galilei group in \cite{BGO92}, where it is showed that $f$ is 
associated to a non--relativistic residue of non--commutativity of boosts 
($[K_1,K_2]=-J$) for Poincar' group, which physically would corresponds to 
a non--relativistic \emph{Thomas precession} \cite{Tho27}. It was also 
noticed that extension $f$ is essentially different from the one 
corresponding to the mass of the system $(m)$. 

However, it seems difficult, from the physical point of view, that such 
\emph{precession} could be a measurable magnitude. We introduce a new perspective 
in favor of this argument; our reasoning is related with the quantization 
process we exposed in previous section, and it departs from the fact that 
if we use a set of suitable canonical coordinates for the orbits with 
$(f,m\neq 0)$ and $(f=0,m\neq 0)$, the parameter $f$ in the first case can 
be embedded in the definition of \emph{position} $\mbf{q}$, but \co 
disappear'' in the expressions of the kernel and trikernel of the SW 
correspondence, which become identical to their analogues for the second 
case. 

This result is in concordance with the ideas exposed in \cite{BGG95} about 
the existence of an isomorphism between all the extended algebras with 
$m\neq 0$ for arbitrary $f$, and give a new point of view to the hypothesis 
that extension linked to $f$ is not physically very relevant in order to 
study the corresponding elementary systems. 

\section*{Acknowledgments}
This work was partially supported by the Direcci¢n General de Ense¤anza 
Superior (DGES, grant PB95-0719) from the Ministerio de Educaci¢n y Cultura 
of Spain and Junta de Castilla y Le¢n.


\end{document}